Active spatial control of terahertz graphene plasmons by tailoring carrier density profile

Ngoc Han Tu[†], Katsumasa Yoshioka, Satoshi Sasaki, Makoto Takamura, Koji Muraki, and Norio Kumada[*]

*NTT Basic Research Laboratories, NTT Corporation, 3-1 Morinosato-Wakamiya, Atsugi, Kanagawa 243-0198, Japan*

**Abstract**

Graphene offers a possibility for actively controlling plasmon confinement and propagation by tailoring its spatial conductivity pattern. However, implementation of this concept has been hampered because uncontrollable plasmon reflection is easily induced by inhomogeneous dielectric environment. In this work, we demonstrate full electrical control of plasmon reflection/transmission at electronic boundaries induced by a zinc-oxide-based dual gate, which is designed to minimize the dielectric modulation. Using Fourier-transform infrared spectroscopy, we show that the plasmon reflection can be varied continuously with the carrier density difference between the adjacent regions. By utilizing this functionality, we show the ability to control size, position, and frequency of plasmon cavities. Our approach can be applied to various types of plasmonic devices, paving the way for implementing a programmable plasmonic circuit.

Controlling spatial patterns of plasmons constitutes a basis for a variety of applications in the fields of plasmonics, metamaterials, and transformation optics[1-3]. Of particular interest is the active control of plasmon confinement and propagation because it will enable us to develop programmable plasmonic circuits. A promising strategy achieving it is to tailor the spatial conductivity pattern in graphene. It has been proposed that plasmonic components such as waveguides, splitters, and switches can be developed in a continuous graphene sheet[4-6]. Using these components, a programmable plasmonic circuit can be configured. Experimentally, however, a platform for implementing this concept has not been developed―most of the intensive works on graphene plasmonics has focused on frequency tuning in a cavity structure with the boundary physically defined by etching[7-10], placing metals[11], or patterning the substrate[12]. While plasmon reflection by an electronic boundary formed at a grain boundary[13,14], moiré-patterned graphene interface[15], and monolayer/bilayer interface[16], or one defined by inhomogeneous chemical doping[17-19] has been observed, these boundaries are unerasable and thus the plasmon reflection cannot be turned off.

To achieve the full control of plasmon reflection/transmission in graphene, independent control of the carrier density on both sides of an electronic boundary is necessary. The optical conductivity of graphene depends on the carrier density n (or Fermi energy $E_\text{F} \propto \pm\sqrt{|n|}$ ), with + for electrons and – for holes) as[20]

$$\sigma = -\frac{ie^2 k_\text{B} T}{\pi \hbar^2 (\omega - i\tau^{-1})} \ln\left[\exp\left(\frac{-E_\text{F}}{k_\text{B} T}\right) + \exp\left(\frac{E_\text{F}}{k_\text{B} T}\right) + 2\right], \tag{1}$$

where $e$ is electron charge, $k_\text{B}$ is Boltzmann's constant, $\omega$ is angular frequency, $\tau$ is scattering time, and $T$ is temperature. Plasmon reflection at a boundary for a step-like $\sigma$ change is given by[21]

$$R = \left|\frac{\sigma_1 - \sigma_2}{\sigma_1 + \sigma_2}\right| \approx \left|\frac{|E_{F,1}| - |E_{F,2}|}{|E_{F,1}| + |E_{F,2}|}\right|. \tag{2}$$

This can be varied between zero and unity by tuning the carrier densities $n_{1,2}$. Therefore, by tailoring the spatial profile of the carrier density, it is possible, in principle, to construct a plasmonic circuit in a continuous graphene sheet.

To control the carrier density profile, electrical gates are ineluctable. Moreover, an electronic boundary has to be sharp to prevent the deterioration of the performance of plasmonic devices, and modulation of the electromagnetic environment must be avoided; otherwise, uncontrollable plasmon reflection is inevitable. The first requirement can be satisfied by placing a patterned gate close to graphene as has been employed for controlling reflection of ballistic electrons at DC regime[22-24]. However, the nearby gate electrode easily affects the electromagnetic environment, so the selection of the gate material satisfying the second requirement is the key issue for the full electrical control of the plasmon reflection. To find an appropriate material, we simulated the coupling in the THz range between uniform graphene and a patterned gate for several conductivity values of the gate electrodes [Fig. 1(a)]. When the gate conductivity is high, acoustic plasmons with the electric field confined between graphene and the gate electrode are formed [inset of Fig. 1(a)][25,26]. This induces unwanted plasmon reflection at the gate boundaries, giving rise to absorption peaks in the transmission spectrum. When the conductivity is lower than $10^3$ S/m, on the other hand, the absorption peaks disappear, and the spectrum can be described by the Drude model of free carrier absorption in uniform graphene. In this case, without having uncontrollable boundary, plasmon reflection can be controlled purely by the carrier density profile.

Based on the simulation, we selected 20-nm-thick zinc oxide (ZnO) with conductivity of ~ 1 S/m for the gate material. To investigate the basic functionality of plasmon reflection represented by Eq. (2), we patterned the ZnO gate into a simple one-dimensional periodic structure with the width of 2 μm spaced by 4 μm. We used a dual-back-gate structure consisting of the ZnO gate and a p-doped semi-transparent Si substrate [Fig. 1(b)]. Monolayer graphene grown by chemical vapor deposition was transferred onto a 21-nm-thick $Al_2O_3$ insulating layer deposited on the ZnO gate. Two ohmic contacts were deposited outside the illumination region. The carrier densities of the graphene on the ZnO gate ($n_{ZnO}$) and on the Si gate ($n_{Si}$) can be independently tuned by the biases on each gate, $V_{ZnO}$ and $V_{Si}$. We measured extinction spectra $1 - T(\omega)/T_{ref}(\omega)$ by Fourier transform infrared spectroscopy with the normally incident radiation polarized perpendicular to the ZnO pattern, where $T_{ref}(\omega)$ is the transmission power at the charge neutrality point (CNP) used as a reference. All the measurements were conducted at room temperature in the vacuum condition.

Figures 2(a) and (b) present the two-terminal resistance obtained by sweeping $V_{ZnO}$ and $V_{Si}$, respectively, while keeping the other gate bias fixed at 0 V. The values of $n_{ZnO}$ and $n_{Si}$ [top axis in Figs. 2(a) and (b)] can be estimated from $n_{ZnO\,(Si)} = C_{ZnO\,(Si)}(V_{ZnO\,(Si)} - V_{ZnO\,(Si)}^{CNP})/e$, where $C_{ZnO\,(Si)}$ is the gate capacitance and $V_{ZnO\,(Si)}^{CNP}$ is the gate bias at the CNP. Using the dielectric constant of 9 and 4 for the $Al_2O_3$ and $SiO_2$ insulating layers, respectively, both $n_{ZnO}$ and $n_{Si}$ at $V_{ZnO} = V_{Si} = 0$ V are calculated to be $\sim -2.7 \times 10^{12}$ cm$^{-2}$, indicating that the graphene is uniformly p-doped at zero gate bias.

Now we know the carrier density profile at each gate bias. To investigate its effects on the plasmon excitations, we measured extinction spectra for several values of $|n_{ZnO}|$ between 0 and $5.4 \times 10^{12}$ cm$^{-2}$ at a fixed $|n_{Si}| = 2.7 \times 10^{12}$ cm$^{-2}$. Figures 2(c)-(e) show the spectra for three representative density profiles. When the carrier density is uniform ($n_{ZnO} = n_{Si}$) [Fig. 2(d)], the extinction increases monotonically with decreasing frequency, as well-described by the Drude model. This verifies that, as simulated, the presence of the ZnO gate does not induce plasmon reflection. Note also that the structural step of ~ 15 nm at the boundaries (supplementary information) does not induce the reflection. When the carrier density modulation is introduced, the incoming light can resonantly excite plasmons with the wavelength determined by the modulation profile. One extreme case is that graphene on the ZnO gate is set to the CNP ($n_{ZnO} \sim 0$). In this case, a single peak appears at 2.3 THz [Fig. 2(e)]. Since plasmons cannot be excited in a system close to the CNP, the observed single peak should be assigned to the plasmon mode confined in the 4-μm-wide graphene regions on the Si gate. Actually, the peak frequency corresponds to the plasmon frequency of 2.3 THz in a graphene micro-ribbon cavity fabricated by etching[7] with the width $W = 4$ μm,

$$\omega_P = \sqrt{\frac{e^2 \hbar v_F \sqrt{\pi |n|}}{2\epsilon^* \epsilon_0 \hbar^2 W}}, \tag{3}$$

where $\epsilon^* = 4$ is the dielectric constant. When $|n_{ZnO}|$ is twice as large as $|n_{Si}|$, two resonance peaks appear [Fig. 2(c)]. The additional peak at 4.7 THz originates from plasmons confined in the 2-μm-wide graphene regions on the ZnO gate. The overall increase in the extinction with decreasing frequency is due to the remnant free carrier absorption, indicating the presence of finite transmission across the boundaries. These results indicate that plasmon reflection can be introduced electrically by the carrier density modulation in continuous graphene.

Next we show the ability to actively control plasmonic cavities in real space. By tuning $V_{ZnO}$ and $V_{Si}$, uniform conditions ($n_{ZnO} = n_{Si}$) at different carrier densities can be obtained [Fig. 3(a)]. Then, by setting one of the gates to the CNP, plasmon cavities are formed. The frequency of the cavity mode [Figs. 3(b) and (c)] depends on the width of the active regions ($W = 2$ or 4 μm) and increases with $|n|^{1/4}$ as expected from Eq. (3) [Fig. 3(d)]. This demonstrates that the spatial position and frequency of a plasmon cavity can be controlled electrically. Real-time switching of plasmon active/inactive regions can be applied for programmable waveguides and switches not limited to cavities.

To extend our device structure to other plasmonic components such as tunable beam splitters and modulators, continuous tuning of the reflection coefficient is necessary. To show how the reflection coefficient evolves with the carrier density profile, we decompose the measured spectra into the free carrier absorption and plasmon resonance contributions by fitting the spectra with the superposition of the Drude function and two Lorentzian functions [Fig. 4(a)]:

$$\sigma(\omega) = \alpha \frac{i}{\omega + i\omega \tau_D^{-1}} + \beta \frac{\omega}{\omega^2 - \omega_{p1}^2 + i\omega \tau_{p1}^{-1}} + \gamma \frac{\omega}{\omega^2 - \omega_{p2}^2 + i\omega \tau_{p2}^{-1}}, \tag{4}$$

where $\alpha$, $\beta$, and $\gamma$ are coefficients. Then, as the measure of the plasmon reflection coefficient, we calculated the area ratio $A_r = A_{Lor}/(A_{Lor} + A_{Dru})$, where $A_{Lor}$ and $A_{Dru}$ are the spectral areas of the Lorentzian and Drude components [pink and sky blue areas in Fig. 4(a)], respectively ($A_r = 0$ and 1 for the perfect transmission and full reflection, respectively). Figure 4(b) shows $A_r$ for different density profiles, $0 \leq |n_{ZnO}| \leq 5.4 \times 10^{12}$ cm$^{-2}$ and $|n_{Si}| = 2.7 \times 10^{12}$ cm$^{-2}$. As discussed above, when the carrier density is uniform ($|n_{ZnO}| = |n_{Si}|$), the spectrum can be described only by the Drude component, which gives $A_r = 0$. As $|n_{ZnO}|$ deviates from $|n_{Si}|$, $A_r$ increases gradually and reaches a maximum of 0.54 at $|n_{ZnO}| \sim 0$. This behavior can be reproduced by $R$ in Eq. (2) [solid line in Fig. 4(b)]. Eq. (2) suggests that the non-perfect reflection at $|n_{ZnO}| \sim 0$ is mostly due to the finite temperature effect (supplementary information). The agreement between the experimental results and the simple model indicates that the reflection coefficient is determined purely by the carrier density profile.

Finally, we discuss the evolution of the resonance frequencies. Figures 4(c) and (d) compile the spectra for $0 \leq |n_{ZnO}| \leq 5.4 \times 10^{12}$ cm$^{-2}$ and $|n_{Si}| = 2.7 \times 10^{12}$ cm$^{-2}$ and their Lorentzian component (the full set of the spectra are presented in the supplementary information). As shown by the peak frequencies as a function of $|n_{ZnO}|$ [Fig. 4(d)], when $|n_{ZnO}| \gtrsim 3.5 \times 10^{12}$ cm$^{-2}$, the spectra show two resonance peaks consistent with the two cavity modes with $W = 2$ and 4 µm [red and blue traces in Fig. 4(e)]. Two resonance peaks also appear for $|n_{ZnO}| \lesssim 1.8 \times 10^{12}$ cm$^{-2}$. Although the frequency of the two independent cavity modes coincide at $|n_{ZnO}| = 0.6 \times 10^{12}$ cm$^{-2}$, the observed peak frequencies show anti-crossing behavior, indicating the presence of capacitive coupling between the two cavity modes. Between the two regions, where $|n_{ZnO}| \sim |n_{Si}|$, the reflection coefficient is not large enough to confine plasmons. In such a regime, the THz radiation can couple plasmons with the wavelength corresponding to the modulation period ($\lambda = 6$ µm). The frequency of this mode is expected to be in the middle between the two cavity modes with $\lambda/2 = 2$ and 4 µm. This model is consistent with the observation of single peaks but not sufficient to fully explain the peak frequencies around $|n_{ZnO}| = 2.7 \times 10^{12}$ cm$^{-2}$. Supper-lattice effects[27] should be involved in this weak reflection regime. Further experiments at lower temperature to obtain sharper resonance peaks would provide a better understanding of the behavior of the resonance frequencies.

In conclusion, we controlled plasmonic response in a continuous graphene sheet by modulating the spatial profile of the carrier density. We demonstrated that position, size, and frequency of plasmon cavities can be controlled electrically. We also showed continuous variation of the plasmon reflection coefficient at a boundary for a step-like change in the carrier density. These functionalities can be applied to a variety of plasmonic applications, not limited to cavities. As it is possible to make the $ZnO/Al_2O_3$ gate multilayer on top and bottom of a graphene sheet, a complex carrier density pattern can be formed. Therefore, our device structure can be a platform for implementing a programmable plasmonic circuit.

**Acknowledgements**

We thank M. Hashisaka and H. Irie for fruitful discussions, M. Ono for valuable help, and A. Tsukada for experimental support.

**Present Addresses**

†RIKEN Center for Emergent Matter Science (CEMS), 2-1 Hirosawa, Wako-shi, Saitama 31-0198, Japan

**Author Contributions**

N.-H. T. fabricated samples and performed all experiments, analyzed data, and wrote the manuscript. K. Y. performed the simulation and discussed the results. S. S. fabricated samples and evaluated the ZnO properties. M. T. grew graphene. K. M. discussed the results. N. K

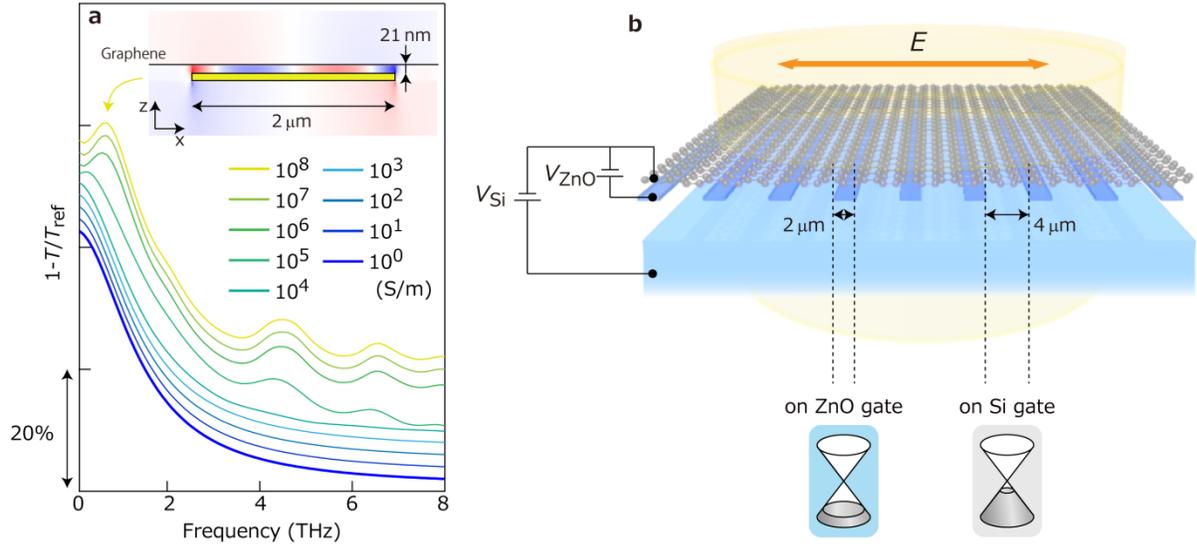

**Figure 1.** Device structure. (a) Simulated transmission spectra for several values of the patterned-gate conductivity. The thickness of the gate was set at 20 nm, two orders of magnitude smaller than the plasmon wavelength in the THz range, to minimize the dielectric modulation. Traces are vertically offset by 2%. Insets show the simulated plasmon field $E_z$ (expanded to the z direction) at 0.7 THz for the patterned-gate conductivity of $10^8$ S/m. The yellow box represents a patterned gate. The conductivity of a typical metal and the ZnO used is $\sim 10^8$ (top trace) and $\sim 1$ S/m (bottom trace), respectively. (b) Schematic representation of the device employing a dual-back-gate structure. The 20-nm-thick ZnO film is patterned into one-dimensional periodic structure with the width of 2 μm spaced by 4 μm on the Si/SiO$_2$ substrate. The patterned ZnO ribbons are connected at both ends to apply the gate bias. The transferred graphene and the ZnO gate are separated by a 21-nm-thick Al$_2$O$_3$ insulating layer. By adjusting gate biases $V_{ZnO}$ and $V_{Si}$, the carrier densities of graphene on the ZnO gate and on the Si gate can be independently tuned. Polarization of the THz light is perpendicular to the carrier density modulation pattern.

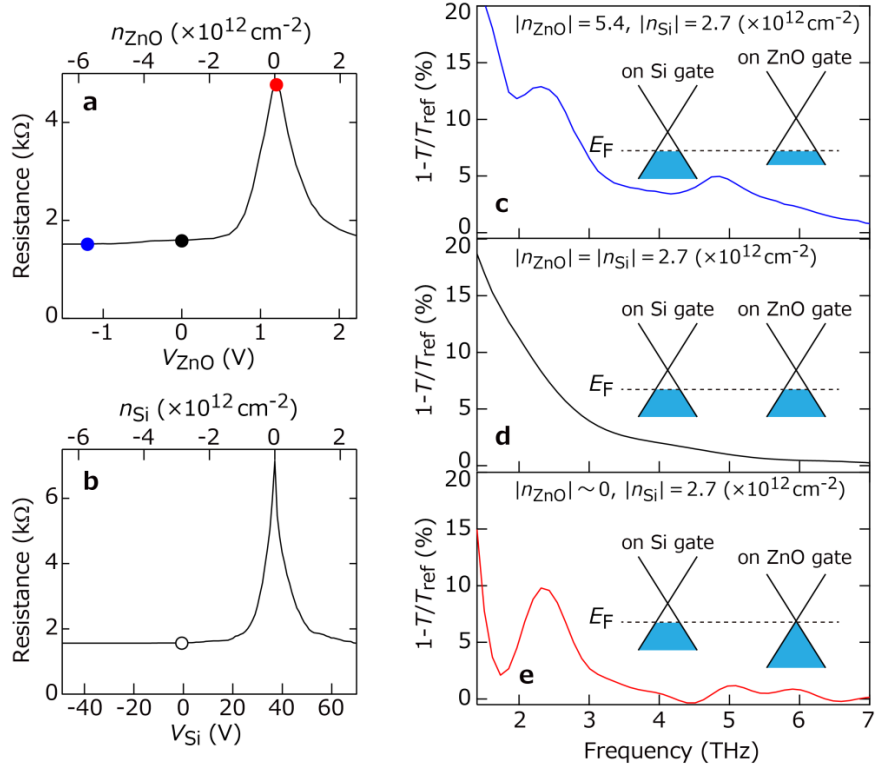

**Figure 2.** Extinction spectra for representative carrier density profiles. (a) and (b) Two-terminal resistance obtained by sweeping $V_{ZnO}$ and $V_{Si}$ while keeping $V_{Si} = 0$ V and $V_{ZnO} = 0$ V, respectively. The current direction is parallel to the ZnO pattern. Carrier density in each region calculated from the gate capacitance and the distance to the CNP is indicated on the top axes. (c)-(e) Extinction spectra $1 - T/T_{ref}$ for three representative carrier density profiles: (c) $|n_{ZnO}| > |n_{Si}| > 0$, (d) $|n_{ZnO}| = |n_{Si}| > 0$, and (e) $|n_{ZnO}| \sim 0$, $|n_{Si}| > 0$. Solid circles in (a) and the open circle in (b) represent the gate conditions used for (c)-(e).

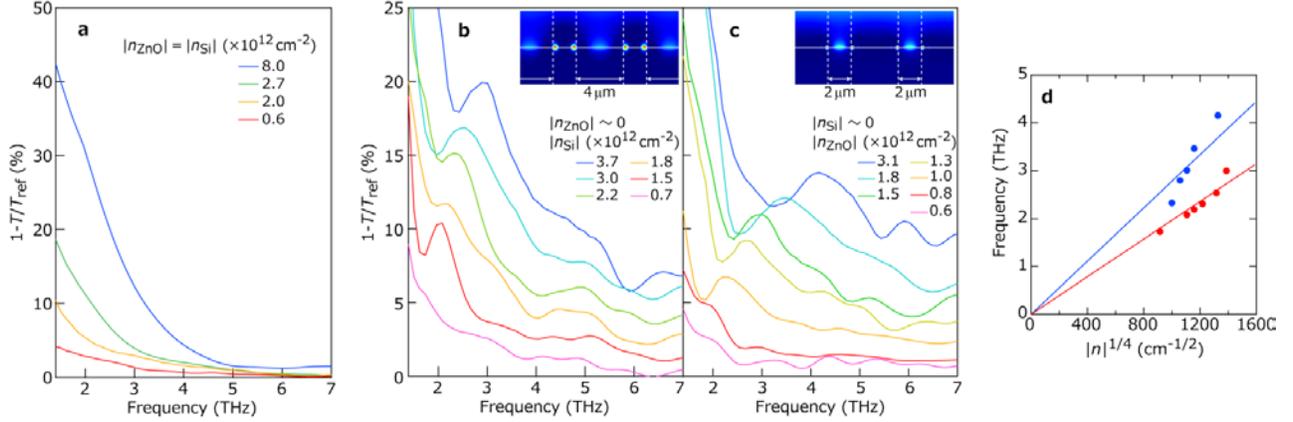

**Figure 3.** Selective activation of plasmonic cavity. (a) Extinction spectra for uniform conditions ($|n_{ZnO}| = |n_{Si}|$) at several values of carrier density. (b) and (c) Extinction spectra for $|n_{ZnO}| \sim 0$ and $|n_{Si}| \sim 0$ for several values of $|n_{Si}|$ and $|n_{ZnO}|$, respectively. Each trace is vertically offset by 1%. The insets show the simulated electric field intensity at resonance frequencies. The horizontal solid and vertical dotted lines indicate the position of the graphene and boundaries, respectively. (d), Plasmon resonance frequency obtained from the spectra in (b) (red circles) and (c) (blue circles) as a function of $|n|^{1/4}$. Red and blue lines represent calculated plasmon frequency in cavities with $W = 4$ and 2 μm, respectively, based on Eq. (3).

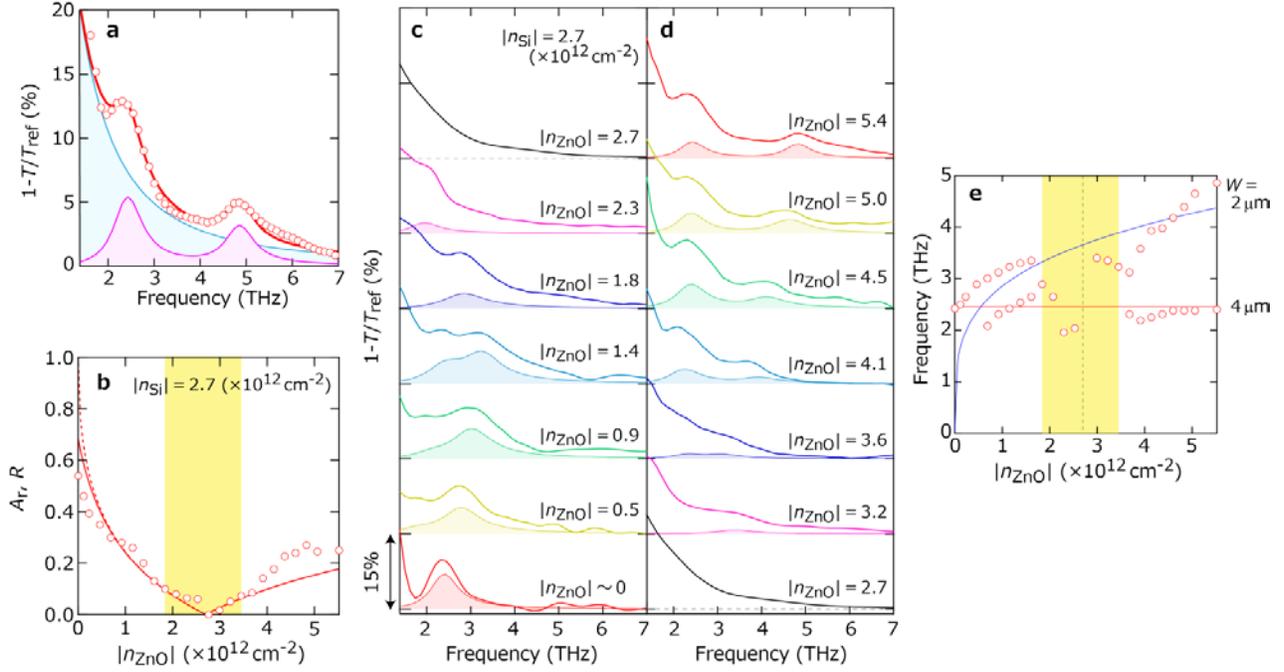

**Figure 4.** Evolution of reflection coefficient and resonance frequency. (a) Spectrum for $|n_{ZnO}| = 5.4$ and $|n_{Si}| = 2.7 \times 10^{12}$ cm$^{-2}$ with fitting line based on Eq. (4). Sky blue and pink areas ($A_{Dru}$ and $A_{Lor}$) represent the Drude and Lorentzian components, respectively. (b) Area ratio $A_r = A_{Lor}/(A_{Lor} + A_{Dru})$ (open circles) as a function of $|n_{ZnO}|$ for fixed $|n_{Si}| = 2.7 \times 10^{12}$ cm$^{-2}$. The solid and dashed lines are the reflection coefficient $R$ obtained from Eq. (2) with $T = 300$ and 0 K, respectively. (c) and (d) Spectra (solid lines) and Lorentzian components (solid areas), where (c) and (d) compile the data for $0 \leq |n_{ZnO}| \leq |n_{Si}|$ and $|n_{Si}| \leq |n_{ZnO}| \leq 5.4 \times 10^{12}$ cm$^{-2}$, respectively. Traces are vertically offset by 15%. (e) Summary of the resonance frequency as a function of $|n_{ZnO}|$. Red and blue lines are the frequency of the cavity modes [Eq. (3)] with $W = 4$ and 2 μm, respectively. The vertical dashed line indicates $|n_{ZnO}| = |n_{Si}|$.

**Supplementary Information**

## 1. Characterization of ZnO as an electrode material for THz experiment

The lack of suitable functional materials for electrodes has limited the practical development of THz devices. Conventional transparent conducting electrodes in the visible range only offer a low transmittance in the THz regime. For instance, even the current industry standard for a transparent electrode material, indium tin oxide (ITO), exhibits high reflection in the THz regime. Here we used a 20-nm-thick ZnO film grown at 180℃ by atomic layer deposition as transparent electrodes in the THz regime. The conductivity of the ZnO film is ∼ 1 S/m. The transmittance of the film is about 90%, and almost independent of the frequency in the whole range of our experiment between 1.3 to 10 THz.

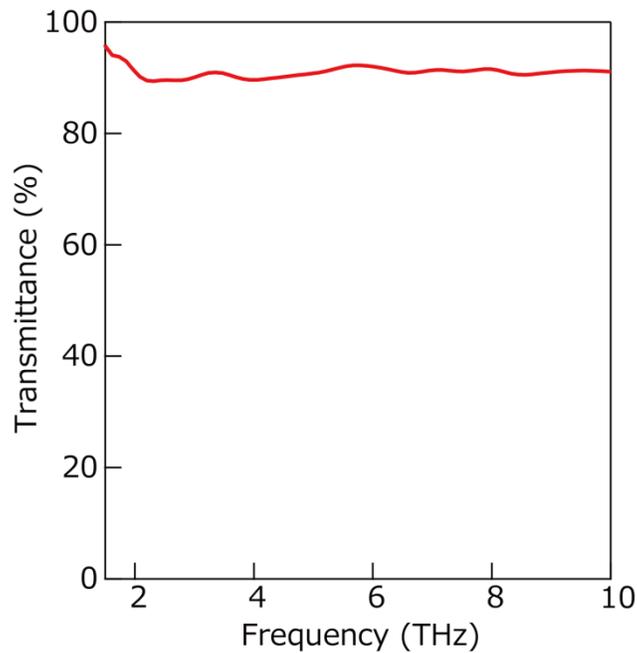

**Figure S1.** Transmittance of the 20-nm-thick ZnO film with the conductivity of ∼ 1 S/m. The contribution of the Si/SiO$_2$ substrate is removed.

## 2. Full set of Extinction spectra for different values of $|n_{ZnO}|$ at a fixed $|n_{Si}|$

Because the space available is limited, we show a limited number of extinction spectra and their Lorentzian components in Figs. 4(c) and (d). Here, we present the full set of the spectra (Fig. S2, left) and the Lorentzian components (Fig. S2, right).

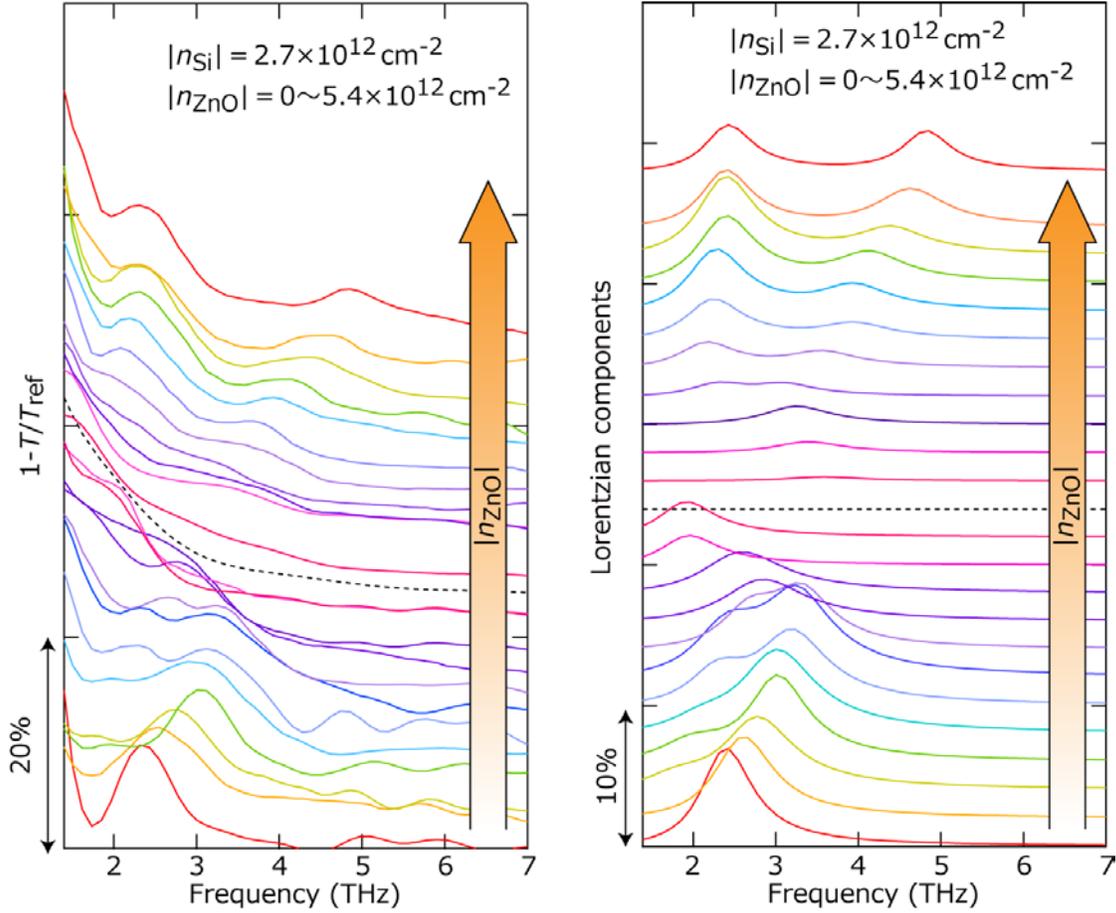

**Figure S2.** Extinction spectra (left) and Lorentzian components (right) for $|n_{ZnO}| = 0 \sim 5.4 \times 10^{12} \text{cm}^{-2}$ and $|n_{Si}| = 2.7 \times 10^{12} \text{cm}^{-2}$.

## 3. Plasmon reflection coefficient for $|n_{Si}| > 0$ and $|n_{ZnO}| \sim 0$

As shown in Fig. 4(b) of the main manuscript, the Drude component is finite even when graphene regions on the ZnO or Si gates are set to the CNP. This is due to the finite conductivity at $|n| \sim 0$ caused by the finite temperature and potential fluctuations $\delta E_F$ induced by disorder. To identify the main cause, we plot $A_r$ for $|n_{ZnO}| \sim 0$ as a function of $|n_{Si}|$ obtained from three samples, A, B, and C, in Fig. S3(b) (sample A was the one used to acquire the data in Figs. 2 and 4, while sample B is the one for Fig. 3). The figure also includes $R$ calculated from Eq. (2), which includes the finite temperature effect, while doesn't include the disorder effect. $A_r$ from sample B and C increases with $|n_{Si}|$, consistent with the tendency of $R$. This indicates that the smaller $A_r$ for smaller $|n_{Si}|$ is due to the finite temperature effect. However, the value of $A_r$ is always smaller than $R$ and depends on the sample. The order of the $A_r$ value, sample A → B → C, agrees with the order of the sample quality, which can be evaluated from the shape of the two-terminal resistance around the CNP [Fig. S3(a)] and the scattering time obtained by the fitting of the spectra [inset of Fig. S3(b)]. In sample A, $A_r$ is about 80% of $R$, suggesting that the main cause the non-perfect reflection ($A_r < 1$) is the temperature effect. In sample B and C, the disorder effect becomes more significant, and the ratio becomes smaller, about 60% and 30%, respectively.

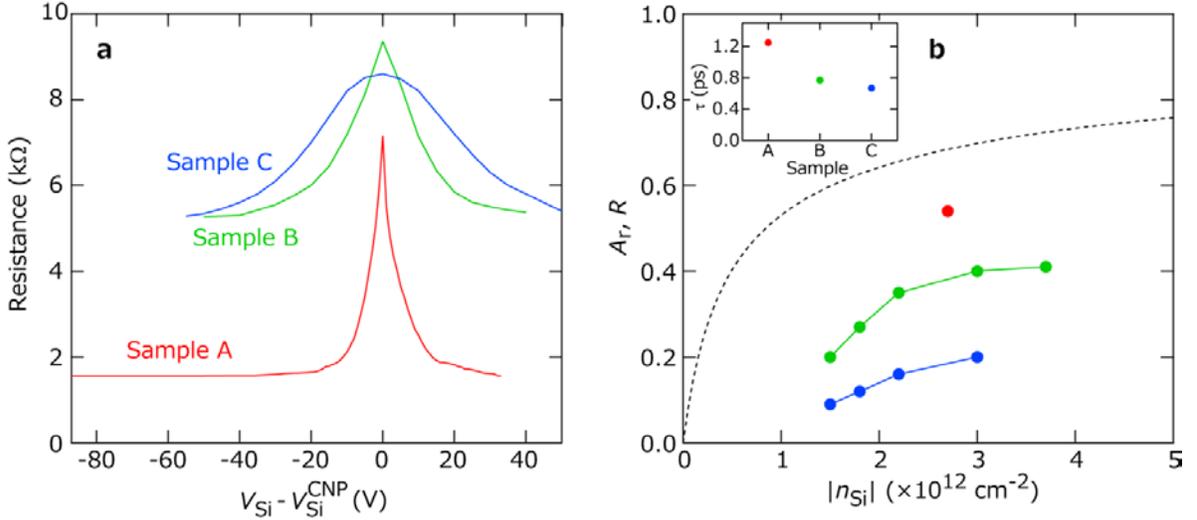

**Figure S3.** (a) Two-terminal resistance for the three samples as a function of $V_{Si} - V_{Si}^{CNP}$. (b) The area ratio $A_r$ (circles) for the three samples and the reflection coefficient $R$ for $T = 300$ K obtained from Eq. (2) (dashed line) as a function of $|n_{Si}|$ at $|n_{ZnO}| \sim 0$. Inset shows $\tau_p$ for $|n_{Si}| \sim 3 \times 10^{12}$ cm$^{-2}$.

## 4. AFM image of the sample

Figure S4 shows an AFM image of our graphene device. Brighter and darker regions correspond to the graphene regions on the ZnO and Si gates, respectively. The step between the two regions is about 15 nm. The AFM image also reveals that monolayer graphene has been uniformly transferred onto the patterned substrate.

Since the step height is two orders of magnitude smaller than the plasmon wavelength, it is reasonable that plasmon reflection is not induced by the step [S1].

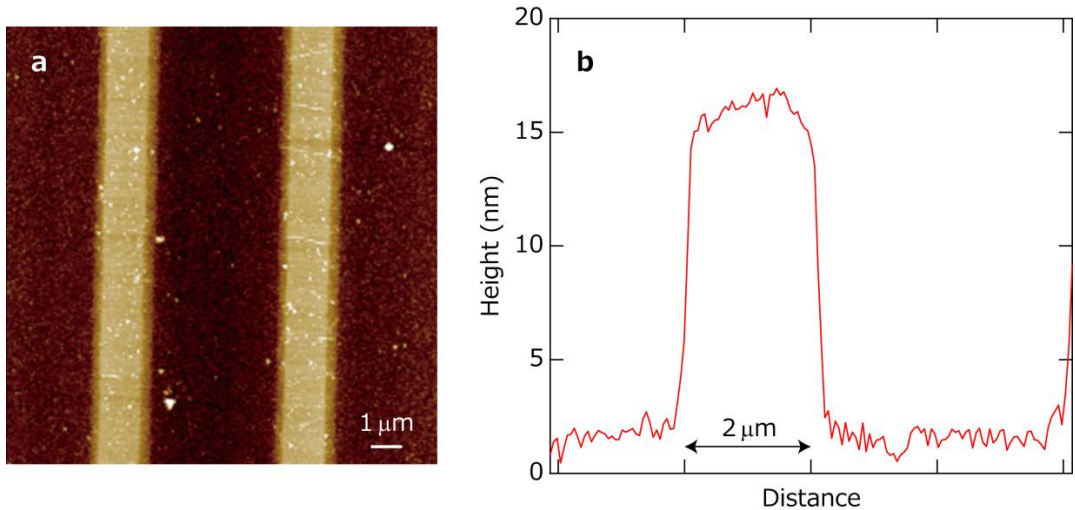

**Figure S4.** (a) AFM image of the graphene device with a patterned ZnO gate. Brighter and darker regions correspond to the graphene regions on the ZnO and Si gates, respectively. (b) Height profile of the same device.

## 5. Electromagnetic simulation of graphene plasmon

A two-dimensional (2D) finite element calculation was performed using commercially available software (COMSOL) to investigate the influence of the gate electrode on the plasmonic response of the continuous graphene. We calculated THz extinction spectra of our device structure, including $Al_2O_3$ insulating layer and $Si/SiO_2$ substrate, for several values of the patterned gate conductivity. The graphene was modeled as a uniform conducting sheet based on the Kubo formula [Eq. (1) of the main text] with $E_F = 0.25$ eV, $\tau = 1$ ps, and $T = 300$ K. The conductivity of the patterned gate was simulated by using the Drude model given by

$$\sigma = \frac{\sigma_{DC}}{1 - i\omega/\Gamma_{gate}},$$

where $\sigma_{DC}$ is the DC conductivity and $\Gamma_{gate}$ is the scattering rate of the gate. The result obtained using the highest conductivity of $10^8$ S/m in Fig. 1b corresponds to the value of the metallic gate (For example, the DC conductivity of gold is ~ $4 \times 10^7$ S/m). On the other hand, the DC conductivity of our ZnO gate is ~ 1 S/m, which is the lowest simulated conductivity in Fig. 1b. We used $\Gamma_{gate} = 50$ THz for every simulation, which is reasonable for the ZnO gate [S2].